\documentclass[aps,prb,twocolumn]{revtex4}

\usepackage{indentfirst}
\usepackage{graphicx}
\usepackage{subfigure}
\usepackage{amsmath}

\begin{document}

\title{Magnetic Small World Nanomaterials: Physical Small World Networks}

\author{M.A. Novotny}
\email{novotny@erc.msstate.edu}
\author{X. Zhang}
\author{J. Yancey}
\author{T. Dubreus}
\author{M.L. Cook}
\author{S.G. Gill}
\author{I.T. Norwood}
\author{A.M. Novotny}
\affiliation{Department of Physics and Astronomy, 
ERC Center for Computational Sciences, 
Mississippi State University, Mississippi State, Mississippi 39762-5167}
\author{G. Korniss}
\email{korniss@rpi.edu}
\affiliation{Department of Physics, Applied Physics, and Astronomy, 
Rensselaer Polytechnic Institute, Troy, NY 12180-3590}

\date{\today}

\begin{abstract}
The question addressed is whether magnetic materials based on physical
small world networks are possible. Physical constraints, such as uniform
bond length and embedding in three dimensions, are the new features added
to make small world networks physical.  Results are presented to further
determine if physical small world networks can exist, and the effect of
the small world connections on the critical phenomena of Ising models on
such networks.  Spectra of the Laplacian on randomly-collapsed bead-chain
networks are studied.  The scaling function for the order parameter of an
Ising model with physical small world connections is presented. 

\end{abstract}

\maketitle

\section{Introduction}

It is well known that novel magnetic behavior is present in systems 
governed by fixed points in other than three dimensions, such as 
quasi-one-dimensional and quasi-two-dimensional 
magnetic systems.\cite{JERO87}  
Recently there has been renewed interest in the behavior of complex 
networks,\cite{ALBE02}  
including small-world (SW) networks.\cite{WATT98}  
These networks mainly originate in
areas in technology, or the social sciences.\cite{NEWM03} 
Properties of networks have
been used in applications, for example, to devise perfectly scalable parallel
algorithms.\cite{KORN03}  
Magnetic models, including Ising 
\cite{GITT00,BARR00,KIM01,HERR02,HONG02,JEON03,HAST03}
models, have been studied on such networks.  
Recently, the question has been asked whether 
or not materials based on physical SW networks 
are possible.\cite{NOVO04} 
Physical constraints, such as uniform bond lengths and node sizes, and that 
the network must be embedded in three dimensions, are the new features we 
add to SW networks to make them physical.\cite{NOVO04}  
Since magnetic models on such SW networks are usually governed by 
a (slightly modified) mean-field fixed point,\cite{HAST03} similar to 
mean-field fixed points for polymer systems,\cite{JANS92,MON93,LUIJ98} 
magnetic spins on physical SW networks should exhibit 
novel magnetic behavior.  

Here we present results to further determine whether physical SW networks
can exist, and the effect of the SW connections on the critical
phenomena of Ising models on such networks.  
We present results of a project with randomly-collapsed bead-chains to 
determine the topological properties and the properties of the Laplacian 
{\em on} such networks. 
The Laplacian operator is the most fundamental operator governing
physical behavior on the network, such as diffusion or collective
excitations in various interacting systems.\cite{MONA99,KOZM04}
The starting lattice for the Ising
simulations are either one-dimensional chains or two-dimensional toruses, with
the physical SW connections added to the underlying lattice.  We
show, for example, how the mean-field behavior of Ising ferromagnets arises
as the lattice size and density of SW connections is varied.  

\section{Magnetic Models}

The magnetic models simulated have Ising spins, $S_i$$=$$\pm1$, 
placed on the 
sites of a one ($d$$=$$1$) or two dimensional ($d$$=$$2$) square lattice, 
with periodic boundary conditions 
and nearest neighbor ferromagnetic exchange coupling $J_1$.  
A selected number of SW bonds, with ferromagnetic 
coupling $J_2$$=$$4J_1$, are also included (Fig.~1).  The Hamiltonian is 
\begin{equation}
{\cal H } = - J_1 \sum_{\langle i,j\rangle} S_i S_j 
- J_2 \sum_{\rm sw} S_{{\rm sw}\>i} S_{{\rm sw}\> j} .
\label{Ising1}
\end{equation}
Our Monte Carlo simulations \cite{LAND00} use a Glauber dynamic 
with random, single-site updates.  When there are 
no SW bonds for the $d$$=$$2$ model the critical temperature 
$k_B T_c \approx 2.269 J_1$.  
A modified Wang-Landau method \cite{WANG01} has been used 
to obtain comparisons with standard Monte Carlo results.  
\begin{figure*}[tb]
\mbox{
 \subfigure
{\includegraphics[height=1.5in]{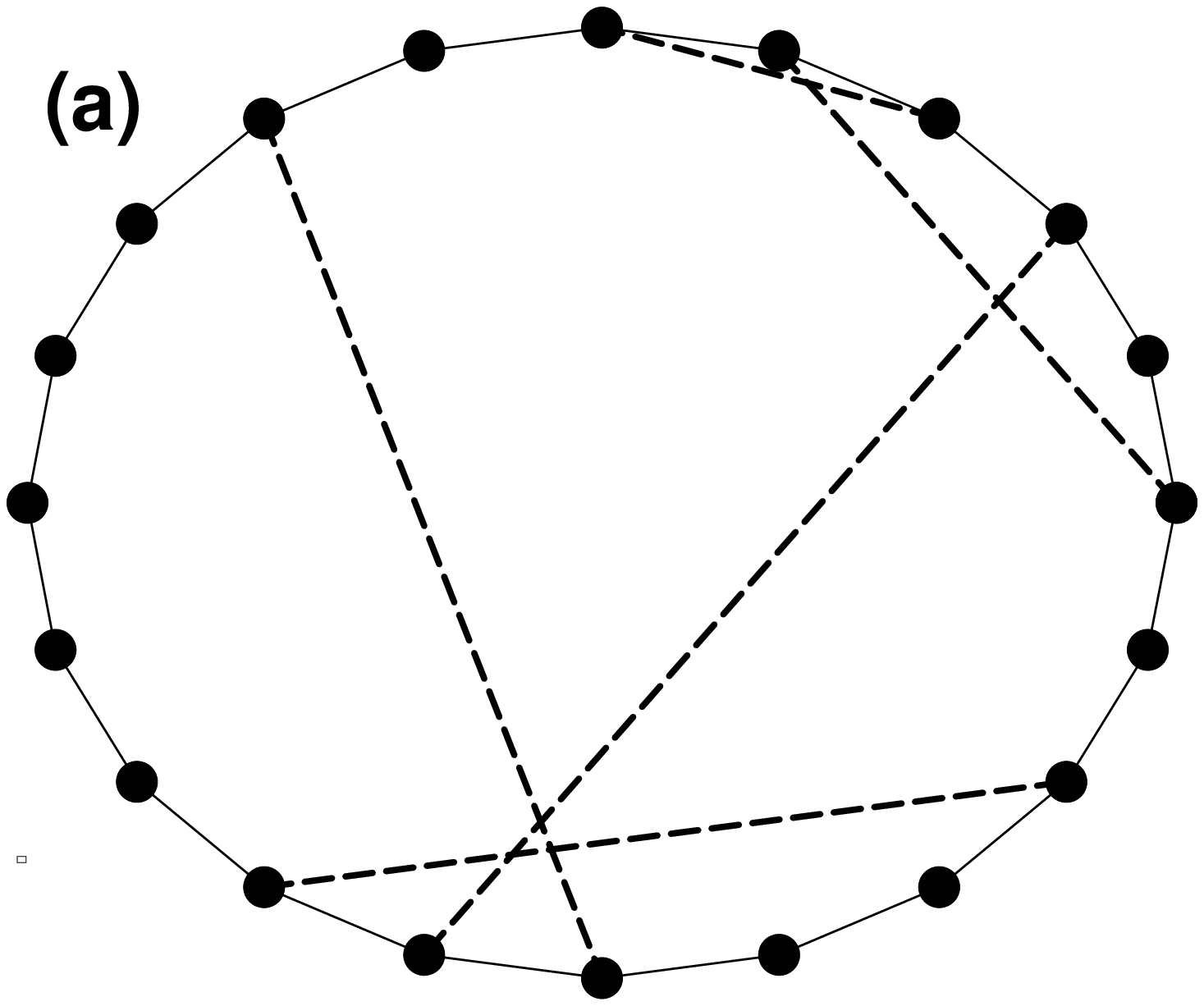}} \qquad 
 \subfigure
{\includegraphics[height=1.5in]{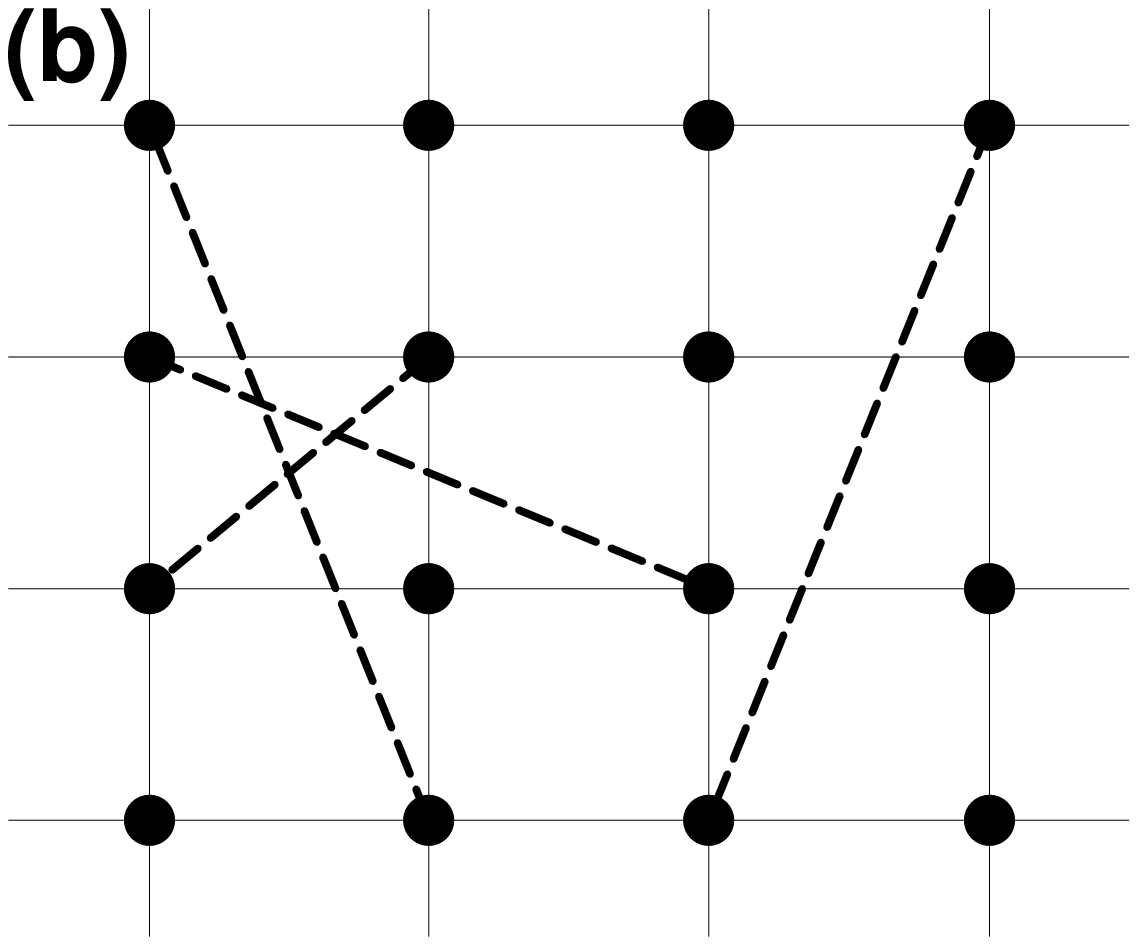}} }
\caption
{\label{FIG1} Examples of the lattices with SW connections 
that were studied.  The filled circles are the 
locations of the Ising spins, the light solid bonds have strength $J_1$ and 
the dashed bonds (the SW bonds) have strength $J_2$.  
(a) A one-dimensional SW graph with $N=20$ and five 
randomly chosen SW bonds.  (b) A square-lattice with 
$L$$=$$4$ with $L$ SW bonds.} 
\end{figure*}

\section{Physical Models}

Two methods to construct physical networks, 
close to SW networks, were used.  
In both cases, the networks formed a physical network because the 
network is embedded in $d$$=$$3$ in such a way that all bond lengths are 
short-ranged and nearly equal.  

In the first method, four high school students 
sequentially numbered all beads, 
randomly collapsed, and glued 
Mardi Gras necklaces of two different lengths.  The bead size was 8~mm, 
and the lengths of the necklaces were $106$~cm 
(about $60$ beads) and $254$~cm (about $155$ beads).  
Beads were cut from the necklaces, with a record kept of the 
adjacency matrix, ${\bf A}$.  The matrix ${\bf A}$ has $1$ as its $i,j$ 
element if bead $i$ is connected 
(either by the necklace string or by glue) to bead 
$j$.  Otherwise the matrix element is zero.  

Physical SW networks were also constructed by implementing a 
simulated annealing program to find an energy minimum of an arrangement of 
atoms starting with a lattice such as in Fig.~1(a).  The energies included 
were those due to the bond lengths, bond angles, dihedral angles, a 
Lennard-Jones 
interaction, and partial ionization.  Energy parameters 
were taken from the AMBER \cite{AMBER} data set.  
These physical SW networks will be analyzed elsewhere, but 
lead to the question of how the number of SW bonds scale with the 
system size.  

\section{Data and Analysis}

The ordered (non-zero) eigenvalues of the (negative) Laplacian on the 
glued Mardi Gras bead-chain networks, 
$- A_{ij} + \delta_{ij}\sum_{\ell}A_{i\ell}$ [Fig.~2], were 
found to be very similar to those of SW networks, such as in Fig.~1(a).
In particular, the value of the smallest eigenvalues (governing, e.g., 
low-energy excitations of such systems) are significantly
larger than those in the spectrum of a pure 1d necklace (where with $N$ 
sites the eigenvalues are $2\left[1-\cos\left(2\pi k/N\right)\right]$ 
with $k=0,1,2,\cdots,N-1$). This 
indicates the opening of a pseudo-gap \cite{MONA99,KOZM04} and the 
corresponding mean-field-like behavior in the physics of fused 1d 
systems, commonly observed in SW networks.\cite{HAST03} 
\begin{figure*}[tb]
\mbox{
 \subfigure
{\includegraphics[angle=0,height=2.0in]{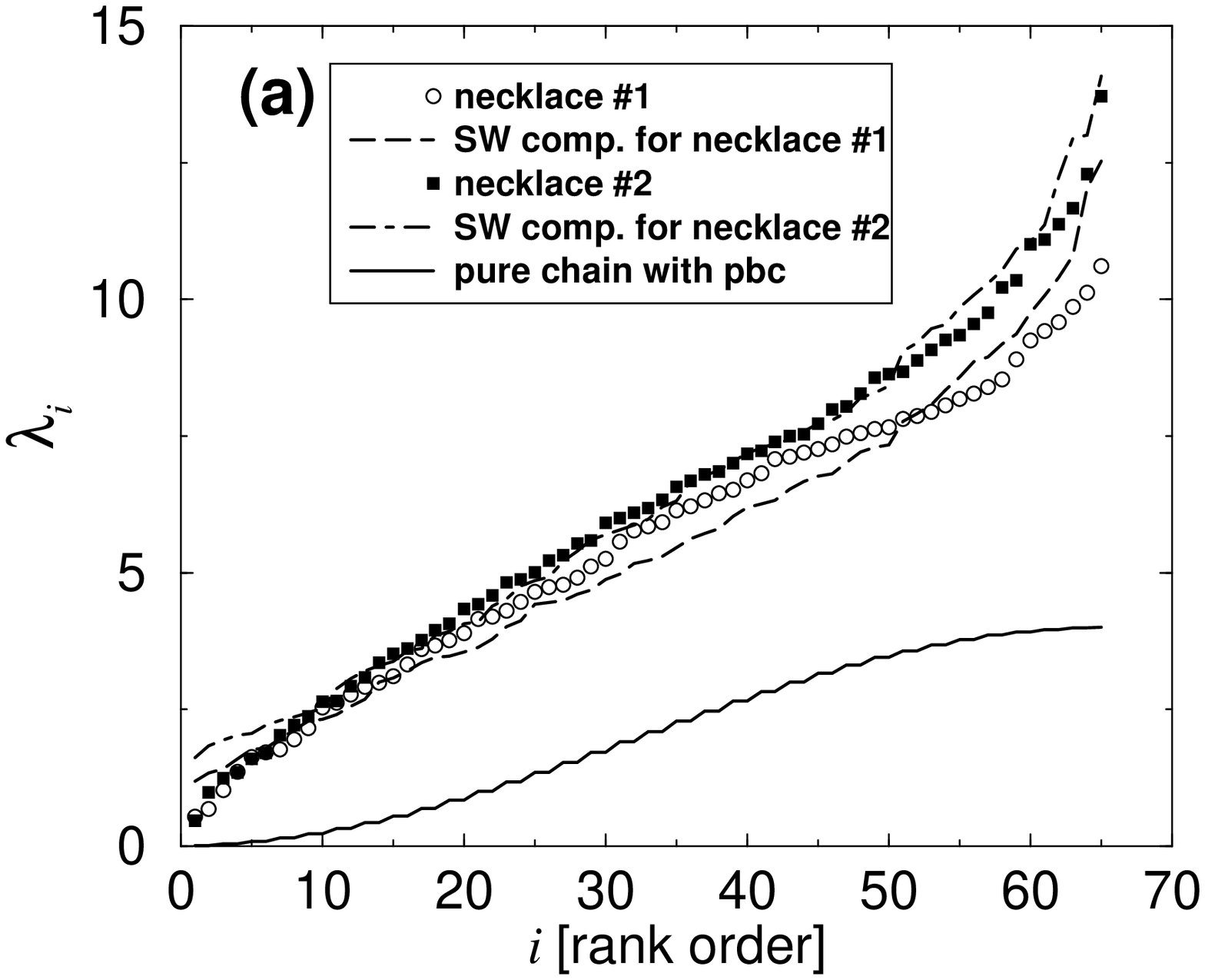}} \qquad
 \subfigure
 {\includegraphics[angle=0,height=2.0in]{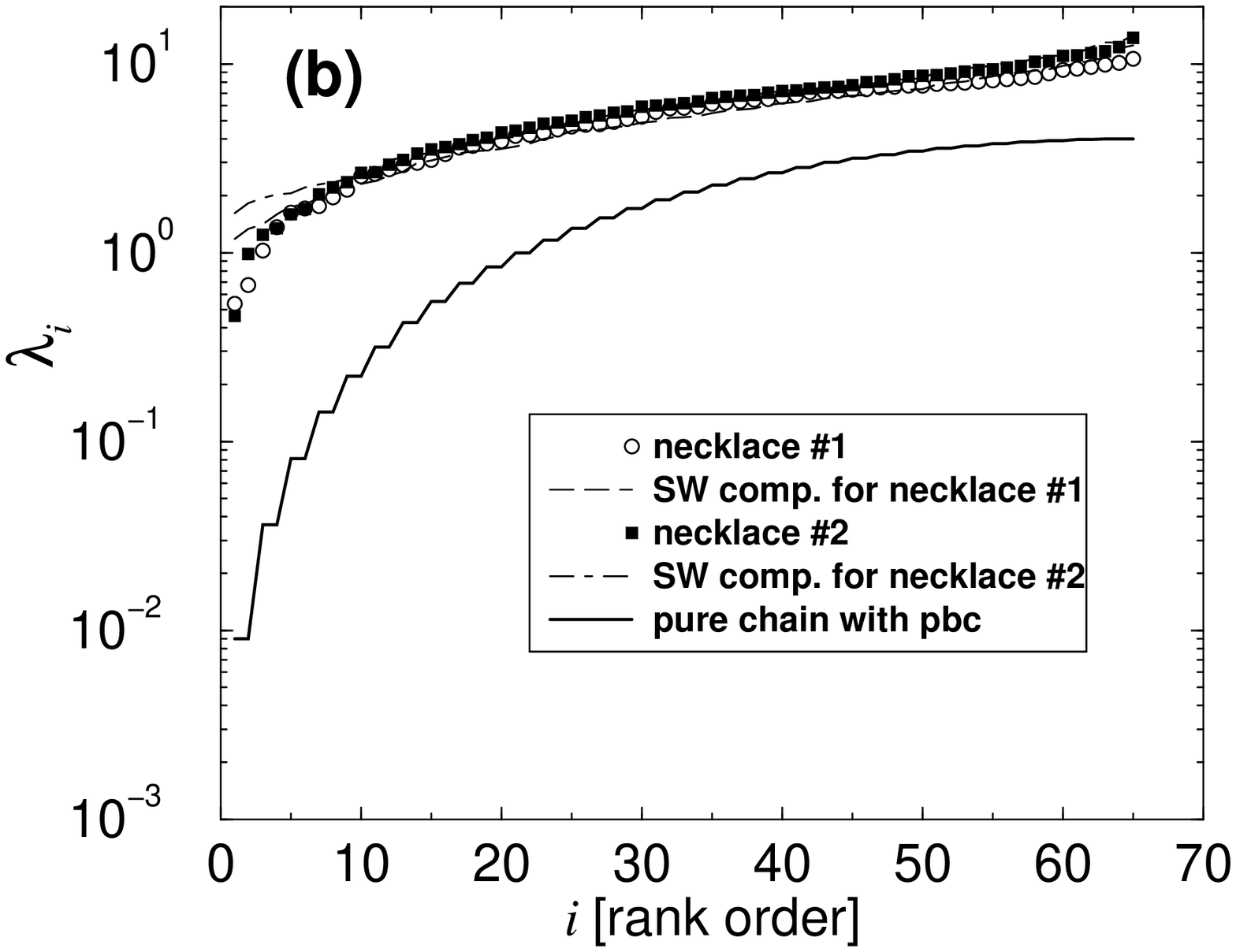}}}
\caption
{ \label{FIG2} The ordered eigenvalue spectrum of the adjacency matrix for 
two different random necklace packings.  
Also shown are corresponding SW realizations as in Fig.~1(a) 
for the same number 
of beads.  Comparison is made to a purely one dimensional lattice.  
Shown are linear (a) and logarithmic (b) scales.} 
\end{figure*}

For physical SW networks, one unanswered question is how 
the number of SW bonds scales with the system size.  Consequently, 
we have investigated cases in which a finite lattice has a reasonable 
number of SW bonds, but the density of SW bonds goes to 
zero as the system size increases.  The order parameter $m$ should 
scale \cite{LAND00} as $m=L^{-\beta/\nu} F(y)$ with 
$y=L^{1/\nu}t=L^{1/\nu}\left|(T-T_{\rm c})/T_{\rm c}\right|$.  
For large $y$ with $L\rightarrow\infty$ and $t\ll 1$ the scaling function 
should asymptotically be $F(y)\approx B y^\beta$ for $T<T_{\rm c}$ and 
$F(y)\approx B' y^{\beta-\nu}$ for $T>T_{\rm c}$.  
Since the density of SW bonds goes to zero 
as $L\rightarrow\infty$, the critical exponents and critical temperature 
should be those of the underlying Ising model.  
Fig.~3(a) shows the scaling function for $m$ for a $L\times L$ 
square lattice with no SW bonds, and Fig.~3(b) for $L$ 
SW bonds of strength $J_2=4J_1$.  Although asymptotically 
the agreement with the normal Ising exponents is apparent, the SW 
bonds modify the scaling function.  This illustrates that some properties of 
the mean-field behavior due to SW 
bonds \cite{GITT00,HAST03,NOVO04} should 
be observed in these finite (hundreds of nanometer in length) systems.  

\begin{figure*}[t]
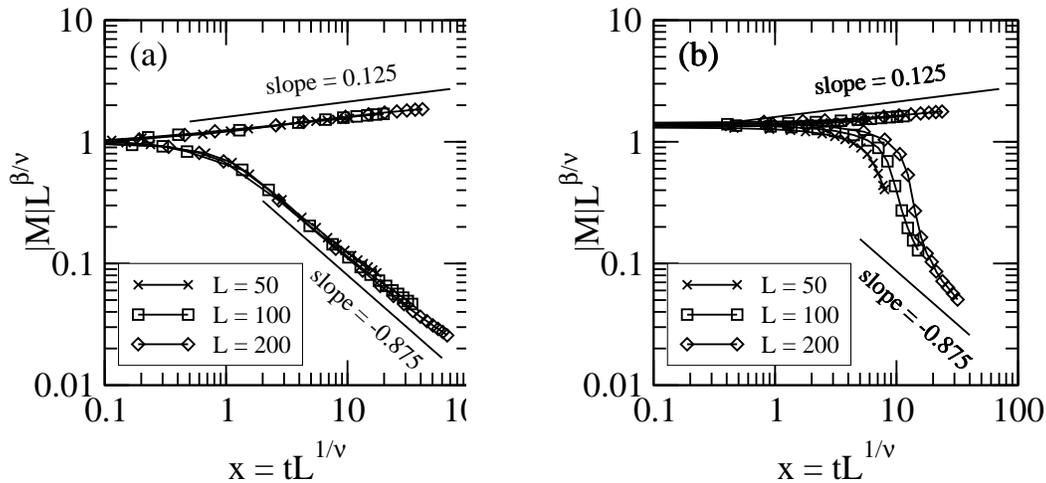

\mbox{
 \subfigure
{\includegraphics[angle=0,height=2.5in]{LSWodp-a.eps}} \qquad
 \subfigure
{\includegraphics[angle=0,height=2.5in]{LSWodp-b.eps}} } 
\caption
{ \label{FIG3} The scaled order parameter for Ising models on $L^2$ 
square lattices.  
The asymptotic slopes have values of $\beta$$=$$\frac{1}{8}$ and 
$\beta-\nu$$=$$-\frac{7}{8}$. (a) for no SW connections, 
(b) for $L$ SW connections, as in Fig.~1(b), with strength $J_2$$=$$4J_1$.} 
\end{figure*}

\section{Conclusion} 

We have presented initial investigations of physical SW networks, and the 
magnetic behavior caused by SW interactions in Ising systems.
Further work is required to definitively illustrate 
the behavior of magnetic systems on physical SW networks.  
Even if physical SW connections are shown to be irrelevant in the 
thermodynamic limit, the scaling functions for nanomaterials change 
substantially, providing quasi- SW behavior.  
Consequently, the current known results should provide a strong impetus 
to attempt to synthesize (or construct) and measure magnetic 
properties of such systems.  

\section{\bf Acknowledgements}
Useful conversations are acknowledged with P.A.\ Rikvold and Z.\ Toroczkai.  
This work was supported by NSF Grants No.\ DMR-0120310, DMR-0113049, 
and DMR-0426488, 
and by the ERC Center for Computational Sciences at 
Mississippi State University.  
The high school students (M.L.C., S.G.G, I.T.N., A.M.N) thank 
T.\ Breckenridge and the ERC for its hospitality.


\begin{thebibliography}{99}


\bibitem{JERO87} {\it Low Dimensional Conductors and Superconductors\/}, 
NATO ASI Series B, Phys.\ Vol.\ 155, editor D.\ Jerome and 
L.G.\ Caron (Plenum, New York, 1987).  

\bibitem{ALBE02} R.\ Albert and A.-L.\ Barab{\'a}si, 
Rev.\ Mod.\ Phys.\ {\bf 74}, 47 (2002).  

\bibitem{WATT98} D.J.\ Watts and S.H.\ Strogatz, 
Nature {\bf 393}, 440, 1998.

\bibitem{NEWM03} M.E.J.\ Newman, 
SIAM Review {\bf 45}, 167, 2003.  

\bibitem{KORN03} 
G.\ Korniss, M.A.\ Novotny, H.\ Guclu, Z.\ Toroczkai, and P.A.\ Rikvold, 
Science {\bf 299}, 677 
(2003).  

\bibitem{GITT00} 
M.\ Gitterman, 
J.\ Phys. A {\bf 33}, 8373 (2000).  

\bibitem{BARR00} 
A.\ Barrat and M.\ Weigt, 
Eur.\ Phys.\ J.\ B {\bf 13}, 547 (2000).

\bibitem{KIM01} B.J.\ Kim, H.\ Hong, P.\ Holme, G.S.\ Jeon, 
P.\ Minnhagen, and M.Y.\ Choi, Phys.\ Rev.\ E {\bf 64}, 056135 (2001).  

\bibitem{HERR02} 
C.P.\ Herrero, 
Phys.\ Rev.\ E {\bf 65}, 066110 (2002).

\bibitem{HONG02} H.\ Hong, B.J.\ Kim, and M.Y.\ Choi, 
Phys.\ Rev.\ E {\bf 66}, 018101 (2002).  

\bibitem{JEON03}
D.\ Jeong, H.\ Hong, B.J.\ Kim, and M.Y.\ Choi, 
Phys.\ Rev.\ E {\bf 68}, 027101 (2003).

\bibitem{HAST03}
M.B.\ Hastings, 
Phys.\ Rev.\ Lett.\ {\bf 91}, 098701 (2003).

\bibitem{NOVO04} 
M.A.\ Novotny and S.M.\ Wheeler,
Brazilian J.\ Phys. {\bf 34}, 395 (2004).

\bibitem{JANS92} S.\ Janssen, D.\ Schwahn, and T.\ Springer, 
Phys.\ Rev.\ Lett.\ {\bf 68}, 3180 (1992).  

\bibitem{MON93} K.K.\ Mon and K.\ Binder, Phys.\ Rev.\ B {\bf 48}, 
2498 (1993).  

\bibitem{LUIJ98} E.\ Luijten and K.\ Binder, Phys.\ Rev.\ E {\bf 58}, 
R4060 (1998).  

\bibitem{MONA99}
R.\ Monasson, Eur.\ Phys.\ J.\ B {\bf 12}, 555 (1999).

\bibitem{KOZM04}
B.\ Kozma, M.B.\ Hastings, G.\ Korniss, Phys.\ Rev.\ Lett.\ 
{\bf 92}, 108701 (2004).

\bibitem{LAND00} D.P.\ Landau and K.\ Binder, 
{\it A Guide to Monte Carlo Simulations in Statistical Physics\/} 
(Cambridge University Press, Cambridge, UK, 2000).  

\bibitem{WANG01} 
F.\ Wang and D.P.\ Landau, 
Phys.\ Rev.\ Lett.\ {\bf 64}, 056101 (2001).  

\bibitem{AMBER} See http://amber.scripps.edu 

\end{thebibliography}
\end{document}